\documentstyle[12pt]{article}
\begin{document}
\topmargin=-.5in
\oddsidemargin=.1in
\evensidemargin=.1in
\vsize=23.5cm
\hsize=16cm
\textheight=23.0cm
\textwidth=16cm
\baselineskip=24pt
\thispagestyle{empty}

\hfill{ITP-SB-97-61}\\
\hfill{hep-th/9712192}\\
\smallskip
\hfill{Revised January 1997}

\vspace{1.0in}

\centerline{\Large \bf FIRST ARE LAST FRACTIONAL-CHARGE SOLITONS}

\vspace{.85in}

{\baselineskip=16pt
\centerline{\large Alfred S. Goldhaber}
\bigskip
\centerline{\it Institute for Theoretical Physics}
\centerline{\it State University of New York}
\centerline{\it Stony Brook, NY 11794-3840}}

\vspace{.5in}

\centerline{\Large Abstract}

\vspace{.5in}

  Jackiw and Rebbi found two types of intrinsically stable or `fundamental'
soliton (kinks in 1+1 D and magnetic monopoles in 3+1 D) which can carry pieces
of elementary-particle charges. After two decades there are no more, and it is
argued here why that is inevitable.

\newpage

Jackiw and Rebbi [JR]'s `Solitons with fermion number 1/2' introduced the 
important concept of fractional soliton charge.    Some definitions 
help
interpret this statement:  

\noindent
1.  In first approximation a soliton may be described as a 
classical field configuration solving nonlinear equations which in turn 
arise from a low-energy, long-distance effective action derived from a 
complete quantum field theory.  

\noindent 
2.  Fractional charges are significant only as eigenvalues 
rather than expectation values; they must be 
absolutely conserved sharp quantum observables.  For
charges to be sharp, in one space dimension 
spatial smearing of the corresponding charge density operator is 
required \cite {KS}, while in higher dimensions temporal smoothing also 
is needed \cite{R,GK}.   

\noindent
3.  One also may ask about fractional charge, ``With respect to what
is it fractional?''  An electron in an insulator has a charge 
$e/\epsilon$, which certainly is a fraction of the charge in free space, 
but this really is an example of charge renormalization rather 
than the phenomenon discussed here.   
Perhaps the most dramatic, well-confirmed example of fractional charge is 
quasiparticle charge in the fractional quantum Hall effect [FQHE], but again the 
fractional value is with respect to that of electron charges measured 
outside the Hall layer, not charges of any particles moving in the layer.  
Remarkable as the quasiparticles  are, they are not charge-fractionating solitons 
\cite{GJ}.  Indeed, it always is consistent to claim that only renormalization
has occurred unless there are objects in some medium whose relative charges
could not be reproduced by finite combinations of quasiparticles, in turn
described as renormalized versions of the particles in another 
medium.  

JR found two types of soliton  which fractionate charge, kinks in one 
space dimension and magnetic monopoles in three space dimensions.   For both 
examples, the key observation is that the spectrum of the Dirac single-particle
Hamiltonian in the presence of the classical background configuration exhibits
a symmetry between positive and negative frequencies, and includes a single
zero-energy bound state.  The empty and full cases for this state should be
charge-conjugate to one another, but differ in fermion charge by $\Delta F=1$,
and hence should carry $F=\pm \frac{1}{2}$.  In
the two decades since, besides the independent work of Su, Schrieffer, and Heeger
[SSH] on kinks in polyacetylene
\cite{SSH}, there
have been a number of studies  confirming and elaborating on the JR
finding.  Shankar and Witten \cite{SW} used bosonization to put
fermions and bosons on the same footing in the kink system, taking account
of possible back reaction by fermion on boson degrees of freedom and
confirming the JR result.   Su and Schrieffer \cite {SS} found examples of 
kinks in
condensed matter models with other rational fractions.  Earlier, Witten
\cite {EW} showed  that magnetic monopoles could have fractional electric charge
determined by the vacuum angle (or equally well by a crossed electric-magnetic
susceptibility like that for a medium with dipolar molecules
carrying both electric and magnetic moments).  Goldstone and Wilczek
\cite{GW} developed an adiabatic flow  analysis (inspiring the approach in
Proof  2 below) demonstrating that the JR monopole belongs to a robust class which
must have fermion charge
$F=\frac{1}{2}$, though not necessarily the fermion zero mode found
by JR.  Callan \cite {CALLAN} considered fermions light
even compared to the Coulomb energy required to confine a unit of electric
charge within a monopole radius, finding $F=\frac{1}{2}$
appears in a natural way.  Over a long period, the notions of
electric-magnetic duality, supersymmetry, and JR fermion zero modes were
locked together,  mutually reinforcing all
three [13-18].   
  However, in all that time no new  types of charge-fractionating soliton
have appeared.  The main purpose of
this work is to explain why there  cannot be any more.  [Such ideas may well be in
the air, for example,  a static electric gauge potential, while it  destroys charge
conjugation symmetry, never produces localized fractional electric charge \cite
{CD}.] 

\bigskip

\noindent
\underline{Whole-Particle Theorem}:   

Only magnetic monopoles 
able to move freely in three space
dimensions, and `kinks'  in one space dimension, can
carry conserved charges which are pieces of those 
carried by weakly-coupled elementary excitations.

\noindent
\underline{Proof 1}:

As long as the elementary charges are perfectly conserved,
only a soliton which in principle could not be created or destroyed
in isolation could carry fractions or pieces of the charges carried
by elementary particles.  [Otherwise, if the soliton could collapse, the fractional
charges would have no possible home.]  We may call such an intrinsically stable
object a `fundamental' soliton. To give precision to this concept, we 
should note that the soliton arises from an effective action reliable at
low energies and long distance scales.  Therefore, the soliton is intrinsically
stable if and only if there is no short-distance, high-frequency modification of
the classical action which would allow creation or destruction (i.e., collapse)
 in
isolation. This means there must be some property of the field configuration at
arbitrarily large distance from the nominal center of the soliton which
unambiguously distinguishes it from the vacuum.  Of course, intrinsic stability
is a necessary condition for charge fractionation, but for any case sufficiency
still must be demonstrated. 

In one space dimension, some field may have a finite or
countable set of allowed values corresponding to degenerate minima in the local
energy density.  If the field has different limits in this `vacuum' class as $|x|
\rightarrow \pm
\infty$,  then infinite action would be required to deform the configuration to a
trivial one, and hence the soliton is absolutely stable -- only
soliton-antisoliton collisions could lead to its disappearance.  Except for such
a `kink', there is no other way to assure absolute stability:  A classical
action with a minimum for some completely localized field
configuration can be modified at short distances in such a way that the minimum
disappears.

In more than one space dimension all fields must go to the same limit in all
spatial directions as distance $r$ from the soliton center increases towards
infinity -- otherwise angular gradients of the fields will lead to energies
diverging with $r$.  There is an exception for gauge theories, where
only the angular {\it gauge-covariant} derivatives must vanish, but by a suitable
choice of gauge the condition of asymptotic angular constancy again may be
imposed.  In that gauge, a vortex in two space dimensions is described by a
 phase jump condition on charged-particle wave functions across some line 
radiating out from
the vortex, such that for the scalar field itself that jump is an integer multiple
of
$2\pi$.
The only other known way to assure absolute conservation is to tie a
charge characterizing the soliton by the Gauss law to a long-range field strength
which at least in principle could be measured at arbitrarily large distances.  In
two space dimensions the only such field strength corresponds to a radial
electric field, but even this has logarithmically infinite energy,
 and therefore cannot be associated with a finite-mass soliton.  In three space
dimensions there are just two possibilities, a radial electric field and a radial
magnetic field.  

Let us first consider the electric case.  If the soliton has a large mass and a
large charge, then by suitable changes in the short-distance dynamics one may
arrange shedding of charge by radiation of light particles, eventually arriving
at a very light object with small charge, which therefore may be treated as an
elementary excitation.  Thus the original heavy object could not possess pieces
of elementary particle charges, as it must be able to decay to a finite number
of elementary particles.  

The only remaining possibility in our world, where electromagnetism is the
unique long-range dynamics coupled to charges which distinguish particles from
antiparticles, is a magnetic monopole in three space dimensions.  Therefore
kinks, vortices, and monopoles are the only fundamental, absolutely stable
solitons, and thus the only ones which might carry fractional charge.  The
magnetic vortex is perhaps the most venerable fundamental soliton, but
we shall see that its ability to carry fractional 
charge is
at best debatable.

\noindent
\underline{Proof 2}:  

Might there be some other
long-range field which could stabilize a soliton, so that the
restricted class found above represents merely a
failure of imagination?  Let us approach the issue
in an independent way, not subject to this question.  As the soliton
must carry a unique, absolutely conserved charge $S$, suppose that
adustments of parameters in the action which do not affect $S$ or its conservation
could decouple the charges $Q_i$ of elementary particles from $S$, so that the
soliton  would have zero values for all of these charges.  If now the parameters
were changed adiabatically to new values which {\it did} engender coupling to
$S$, then the only way the soliton could acquire fractional charge values would
be by adiabatic current flow in from (or out to) infinity.  In a region of space
far from the soliton, imagine that some parameter $M$ has a small logarithmic time
derivative, $$d\ln M/dt = \Lambda.$$  We now require a current $${\bf J} =  
\Lambda {\bf K}$$ to carry the charge in from infinity. In one space dimension
only a sign is needed to specify current direction, so to have the right parity it
is sufficient for 
$\Lambda$ to be a pseudoscalar -- no explicit ${\bf K}$ is needed.  In more than
one space dimension
$n$, current conservation implies ${\bf K} \propto {\bf r} \ 
r^{-n}.$  As this becomes arbitrarily small at large distances, 
${\bf K(r)}$ should be obtainable by perturbation theory in terms of observable
fields emanating from the soliton.   Leaving gravity aside, up to $n=3$
the only such fields
allowed by known low-energy physics are those of electric or magnetic monopoles.

For electric monopoles, the adiabatic current simply gives the standard
renormalization group flow of electric charge.  This applies to all objects,
whether elementary particles or solitons, and therefore cannot be an
illustration of fractional soliton charge.    

For magnetic monopoles, if $\Lambda$ is a pseudoscalar (as in the kink
example) there will be adiabatic flow of suitable charges, such as 
electric charge \cite{EW} or fermion
charge \cite{GW}.  Thus the assumption
  that the conserved soliton charge is not intrinsically correlated with
conserved elementary particle charges is sufficient to single out
kinks and monopoles as the only possible charge-fractionating solitons.

\noindent
\underline{Comments}:

1) As discussed earlier, the quasiparticles of the
fractional quantum Hall effect carry a fraction of the
 electron charge in other media.
Laughlin
 \cite {Laugh} used adiabatic flow considerations to show this:  Imagine an
arbitrarily thin tube of magnetic flux which pierces the Hall surface.  As the
magnitude of the flux increases from zero to one quantum, Faraday's law requires
an azimuthal electric field, and therefore a  Hall current in from infinity which
carries a total charge
$\nu e$, where $\nu$ is the Hall fraction.  Now for any integer number of flux
quanta the Hamiltonian  is identical to that for zero flux, so if the system
initially were in its ground state this flow must have produced a localized
excitation with charge which is a fraction of $e$, and whether the excitation
represents one or several quasiparticles, the charge of a quasiparticle therefore
is fractional.  The
quasiparticles clearly are not solitons, and there are no larger-charge, stable
particles in the Hall medium, so in no sense are the quasiparticles an example
of the JR phenomenon.  The fact that they have finite mass results because they
are hybrids between two and three space dimensions, so that their Coulomb energy
is finite, unlike that of fully (2+1 D) objects.

2) The concept of fractional soliton charge may be traced to Skyrme \cite
{THRS}, who argued that his classical field configuration could be quantized with
half-integer isospin and spin (a possibility shown consistent
with the usual spin-statistics connection in \cite {FR}).
 Half-integer values allow the skyrmion to be identified 
with the nucleon, but by the whole-particle theorem are impossible without
fundamental isospinor fermions.  Microscopic
analyses agree,
indicating that the spin and isospin of the skyrmion will be integer or
half-integer as the number of  colors of up and down quarks is even or odd \cite
{WF}.  Lesson:  Fundamental solitons must be topological, but
topological solitons may not be fundamental.  

 Skyrme's model describes the nucleon
entirely in terms of an $SU(2)$ matrix function
$U({\bf r},t)$, while in a `hybrid'
model the $U$ function is used outside a chosen `bag' radius $R$,
and inside are free quarks with angle-dependent
boundary condition parametrized by $U({\bf r})$ at the bag wall
\cite {RGB}.  
Goldstone
and Jaffe \cite{GJ'} showed that the simple
boundary condition guessed in \cite{RGB} meets the requirement of net
integer baryon number $B$.

3) The SSH kink analysis \cite{SSH} shows that in one space dimension `spinons'
with spin $\frac{1}{2}$ but no charge and `holons' with charge $\pm e$ but no
spin can travel independently.  Kivelson et al. \cite{KRS} proposed that such
objects might play a role in the planar dynamics which appears to be critical in
high $T_C$ superconductivity.  Detailed studies suggest that if so, either these
fractional objects are connected by strings \cite{L'} or are able to move only
along particular lines in the plane \cite{EK}, in agreement with the
whole-particle theorem. 

4) The uniqueness of magnetic monopoles as possible carriers of fractional
particle charge is connected with other special properties, such as their ability
to convert the dynamics of the lowest fermion partial wave into a one-dimensional
problem on a half-line.  This is an example of the fact that the chiral 
anomaly in three space dimensions may be written as the product of
 a magnetic-field
contribution which reduces the problem to one space dimension, and an 
electric-field contribution just like that for QED in one space
dimension \cite {AGP}.  The same long-range magnetic field is responsible
for the unique possibility of creating a fermion from bosons in a
world with no fundamental fermons \cite{ASG0}.

5) The assertion that only magnetic monopoles can carry
fractional soliton charge in three space dimensions seems to violate
electric-magnetic duality.  However, our world has
no light, elementary magnetic monopoles.
If instead there were light, weakly-coupled monopoles, objects with
(inevitably) large electric charge could carry fractions of the charges of
the elementary poles, clearly fulfillng duality.

6)  Proof 2 is tested by Higgs-Chern-Simons vortices in abelian 2+1 D gauge
theory.  There is a locally conserved charge $Q=\kappa\Phi + q_H$, with $\kappa$
the Chern-Simons coupling, $\Phi$ the quantized magnetic flux, and $q_H$ the
Noether charge of the Higgs field, not separately conserved. 
As the source of an electric field strength which must vanish exponentially at
large radius, $Q$ vanishes by the Gauss law.  Indeed, with the gauge kinetic term
$F^2$ omitted, the resulting `self-dual' vortex \cite {JLW} has vanishing $Q$
density everywhere!  This system manifestly violates electric charge
conjugation symmetry, but does not generate fractional values for the conserved
charge. 

In sum, the existence even in principle of an adiabatic path for breakup of
conserved elementary particle charges, natural whenever solitons may be
treated as approximately classical,
 is enough to assure that only kinks or monopoles can nucleate such breakup.

A sceptical remark of Jeffrey Goldstone about skyrmions
with fractional $B$ stimulated this study, supported in part by
the National Science Foundation.  Roman Jackiw, Jainendra Jain, Vladimir Korepin,
and Misha Stephanov gave instructive comments.

\end{document}